\documentclass[a4paper,11pt,pdftex]{article}

\usepackage[utf8]{inputenc}

\usepackage[dvips]{graphicx}
\DeclareGraphicsExtensions{.pdf, .jpg, .png} 
\graphicspath{{./fig/}}

\usepackage[section]{placeins} 

\usepackage{pslatex}

\usepackage{geometry}

\usepackage{array}
\newcolumntype{A}{>{\centering\let\newline\\\arraybackslash\hspace{0pt}}m{0.26\textwidth}}
\newcolumntype{B}{>{\centering\let\newline\\\arraybackslash\hspace{0pt}}m{0.18\textwidth}}

\usepackage[normalem]{ulem}
\usepackage{ifthen}
\usepackage{comment}

\usepackage{bigints}
\usepackage{scrextend}

\usepackage{xcolor}
\usepackage{soul}
\usepackage{epsfig}
\usepackage[numbers,sort&compress]{natbib} 

\usepackage[font=footnotesize,hang]{caption} 
\usepackage{euscript}
\usepackage[bottom]{footmisc}
\usepackage{gensymb} 
\usepackage{xspace} 
\usepackage[USenglish]{babel} 
\usepackage{microtype}

\usepackage{lineno}

\usepackage[T1]{fontenc}

\usepackage{verbatim}
\usepackage{numprint} 
\usepackage{eso-pic}
\usepackage{subfigure}
\usepackage{multirow}
\usepackage{epigraph}

\usepackage{authblk}

\usepackage[useregional]{datetime2}

\usepackage{fancyhdr}
\usepackage{lastpage}
\DeclareSymbolFont{Letters}{U}{zeur}{m}{n} 
\DeclareMathSymbol{\psi}{\mathalpha}{Letters}{"20} 

\usepackage{amsfonts}
\usepackage{amssymb}
\usepackage{amsmath}
\usepackage{mathrsfs} 

\usepackage{hyperref} 

\let\phi\varphi
\let\theta\vartheta

\renewcommand{\exp}{\operatorname{e}^} 

\renewcommand{\Re}{\operatorname{Re}}

\renewcommand{\imath}{\operatorname{i}}

\title{Retrieval of phase relation and emission profile of quantum cascade laser frequency combs} 

\date{}

\author[1,*,$\dagger$]{Francesco Cappelli}
\author[1,*,$\dagger$]{Luigi Consolino}
\author[1]{Giulio Campo}
\author[1,2]{Iacopo Galli}
\author[1,2]{Davide Mazzotti}
\author[1]{Annamaria Campa}

\author[3]{Mario Siciliani de Cumis}

\author[1,2]{Pablo Cancio Pastor}
\author[1]{Roberto Eramo}

\author[4]{Markus Rösch}
\author[4]{Mattias Beck}
\author[4]{Giacomo Scalari}
\author[4]{J\'er\^ome Faist}

\author[1]{Paolo De Natale}
\author[1,2]{Saverio Bartalini}

\affil[1]{CNR-INO -- Istituto Nazionale di Ottica, Largo Enrico Fermi 6, 50125 Firenze FI, Italy  \newline \& LENS -- European Laboratory for Non-Linear Spectroscopy, Via Nello Carrara 1, 50019 Sesto Fiorentino FI, Italy}
\affil[2]{ppqSense Srl, Via Gattinella 20, 50013 Campi Bisenzio FI, Italy}
\affil[3]{ASI -- Agenzia Spaziale Italiana, Contrada Terlecchia, 75100 Matera MT, Italy}

\affil[4]{Institute for Quantum Electronics, ETH Zurich, 8093 Z\"urich, Switzerland}

\affil[*]{These two authors contributed equally to this work.} 
\affil[$\dagger$]{Corresponding authors: \texttt{francesco.cappelli@ino.it}, \texttt{luigi.consolino@ino.it}}

\begin{document}

\maketitle
\thispagestyle{fancy}

\begin{abstract}

The major development recently undergone by quantum cascade lasers has effectively extended frequency comb emission to longer-wavelength spectral regions, i.e. the mid and far infrared. Unlike classical pulsed frequency combs, their mode-locking mechanism relies on four-wave mixing nonlinear processes, with a temporal intensity profile different from conventional short-pulses trains. Measuring the absolute phase pattern of the modes in these combs enables a thorough characterization of the onset of mode-locking in absence of short-pulses emission, as well as of the coherence properties. 

Here, by combining dual-comb multi-heterodyne detection with Fourier-transform analysis, we show how to simultaneously acquire and monitor over a wide range of timescales the phase pattern of a generic frequency comb. The technique is applied to characterize a mid-infrared and a terahertz quantum cascade laser frequency comb, conclusively proving the high degree of coherence and the remarkable long-term stability of these sources. Moreover, the technique allows also the reconstruction of electric field, intensity profile and instantaneous frequency of the emission. 

\end{abstract}

\section{Introduction}

The optical frequency comb (FC) is a peculiar multi-frequency coherent photonic state made of a series of evenly-spaced modes in the frequency domain, typically generated by frequency-stabilized and controlled femtoseconds mode-locked lasers \cite{Jones:2000,Diddams:2000,Holzwarth:2000,Udem:2002}. In the visible and near infrared (IR) such technology is nowadays well established \cite{Diddams:2010}. The miniaturization of these sources, together with the expansion of their operation range towards other spectral regions (e.g. mid and far IR), is crucial for broadening their application range. 

In this direction, the most interesting results have recently been achieved with quantum cascade lasers (QCLs), current-driven semiconductor lasers based on intersubband transitions in quantum wells, emitting high-power coherent radiation in the mid IR and terahertz (THz) \cite{Faist:1994,Beck:2002,Kohler:2002,Tombez:2013a,Consolino:2018a,Faist:2013bo}. In these devices the upper lasing state lifetime is very short compared to the cavity round-trip time. Therefore, energy cannot be stored during the round trip, the formation of optical pulses is prevented and classical passive mode locking is not achievable \cite{Hugi:2012,Malara:2013,Faist:2016}. 

Recently, active pulsed mode locking has been achieved both in mid-IR \cite{Wang:2009,Revin:2016} and THz \cite{Barbieri:2011,Wang:2017} QCLs. The limitation of this approach derives from the need of close-to-threshold operation in order to mitigate gain saturation, significantly limiting the emitted power, and from the length of the pulses that cannot reach the inverse of the gain bandwidth.

Quite astonishingly, by using broadband Fabry-P\'erot QCLs \cite{Riedi:2013,Riedi:2015} designed to have a low group velocity dispersion, FC generation was demonstrated in fully free-running operation (QCL-combs) \cite{Hugi:2012,Burghoff:2014,Rosch:2014}. Starting from the independent longitudinal modes generated by a Fabry-P\'erot multimode laser, degenerate and non-degenerate four-wave mixing (FWM) processes induce a proliferation of modes over the entire laser emission spectrum \cite{Friedli:2013}. The original modes are then injection-locked by the modes generated by FWM, ensuring a fixed phase relation among all the longitudinal modes, giving birth to a FC \cite{Faist:2016}. 

Various techniques to characterize the emission of mid- and far-infrared QCL-combs have been developed. Actively mode-locked QCL-combs can, in principle, be characterized by means of techniques based on second harmonic generation, such as frequency-resolved optical gating (FROG) \cite{DeLong:1994,Freeman:2013}. On the other hand, FWM-based QCL-combs have been characterized with techniques waiving nonlinear effects \cite{Hugi:2012,Burghoff:2014,Villares:2014,Cappelli:2016,Cappelli:2015,Burghoff:2015,Singleton:2018}. Among them, only the shifted-wave interference Fourier-transform spectroscopy (SWIFTS) technique \cite{Burghoff:2015,Singleton:2018} can access the phase domain by measuring the phase difference between adjacent FC modes, and therefore retrieving the phase relation of continuous portions of the FC spectrum by a cumulative sum on the phases. This latter is one of the main limitations of the SWIFTS technique, together with the need of a mechanical scan, not allowing a simultaneous analysis of the phases. 

In this paper, we propose and experimentally demonstrate the possibility of acquiring in real-time the Fourier phases of the modes of a generic FC. The proposed approach, based on a dual-comb multi-heterodyne detection and a subsequent Fourier transform analysis, allows both a simultaneous retrieval of the modes phases of FCs with spectra of any shape, and monitoring the evolution of the phase relation. We call this method Fourier-transform Analysis of Comb Emission (FACE) and we consider it as the most direct and general method for a thorough FCs characterization, with the additional advantage to be suitable for any spectral region of interest. 

A mid-IR and a THz QCL-comb have been characterized, showing both robust correlation among the modes over a ms timescale, and remarkable stability over tens of minutes. For the THz QCL-comb the complete set of measured phases and amplitudes allows to reconstruct the emitted intensity profile and the instantaneous frequency, confirming 
a hybrid frequency/amplitude modulation regime in the device emission.



\section{Results}

\subsection{Measurement technique}

A FC is made of a series of evenly-spaced modes in the frequency domain. The overall emitted electric field can be written as: 
\begin{equation}
E_\text{FC} = \sum_n E_n \exp{\imath [ 2 \pi (n f_\text{s} + f_\text{o}) t + \phi_n ] } 
\label{eq:FCfield}
\end{equation}
Here $n$ is the integer numbering the $n$-th FC mode, $E_n$ is the (real) amplitude of the $n$-th mode, the exponent is the total phase of the $n$-th mode, $f_\text{s}$ is the mode spacing, $f_\text{o}$ is the offset frequency, $t$ the time, and $\phi_n$ is the Fourier phase of the $n$-th mode. The condition for $E_\text{FC}$ representing a FC field is that all the $\phi_n$s are constant in time, with fluctuations $\delta \phi_n \ll \pi$, meaning that a well-defined phase relation among the modes is established, without any constraint on the actual shape of the phase relation itself.

\begin{figure}[!htbp]
\begin{center}
\includegraphics[width=0.9\textwidth]{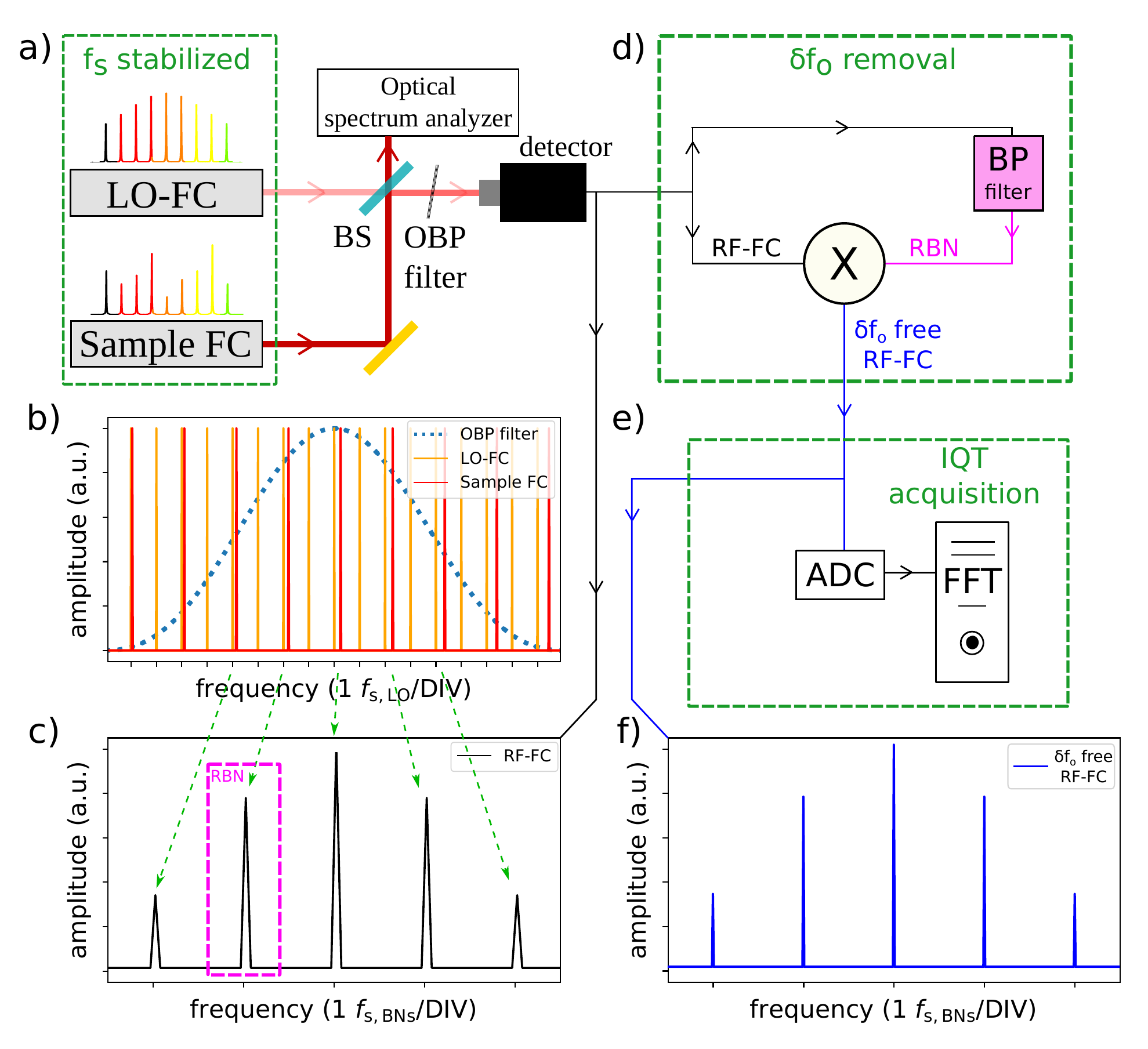}
\caption[]{Experimental setup. \newline a) Dual-comb multi-heterodyne detection scheme. LO: local oscillator, reference FC. BS: beam splitter. OBP: optical band-pass filter. b) Sample FC spectrum, LO-FC spectrum and optical band-pass filter. c) Representation of the RF-FC spectrum with the selection of the reference beat note (RBN). d) Electronic setup used for common-mode noise subtraction. BP filter: electronic band-pass filter. e) Quadratures acquisition of the signal time traces and FFT operation. ADC: analog-to-digital converter (digitiser). FFT: fast Fourier transform routine. f) Common-mode-noise-free RF-FC spectrum. } 
\label{fig:dual-comb_setup}
\end{center}
\end{figure}

The information on the Fourier phases of the FC modes is stored in the time-varying electric field of the emitted radiation and can be extracted by a fast Fourier transform (FFT) analysis. In order to be acquired, the field oscillating at optical frequencies ($> 1$~THz) is down-converted to the radio frequencies (RF) by beating the sample FC with a second local oscillator FC (LO-FC). This  dual-comb multi-heterodyne detection scheme \cite{Keilmann:2004,Coddington:2016}, sketched in fig.~\ref{fig:dual-comb_setup}a, allows a one-by-one mapping of the sample FC in the RF (RF-FC, see fig.~\ref{fig:dual-comb_setup}b and c), with a spacing between the beat notes (BNs) 
\begin{equation} 
f_\text{s,BNs} = | f_\text{s,sample} - k \cdot f_\text{s,LO}|
\label{eq:RFspacing}
\end{equation}
being $f_\text{s,sample} > f_\text{s,LO}$ and $k$ the integer minimizing $f_\text{s,BNs}$. 
Naming $\phi_{n\text{,sample}}$ and $\phi_{m\text{,LO}}$ the Fourier phases of the sample FC and of the LO-FC, respectively, the phases of the related BNs are given by $\phi_{i\text{,BN}} = \phi_{n\text{,sample}} - \phi_{m\text{,LO}}$ (see section \ref{sec:BNs_phases}). By measuring $\phi_{i\text{,BN}}$, the Fourier phases of the sample FC are determined against those of the LO-FC. 

Differently from dual-comb spectroscopy (DCS) \cite{Coddington:2016}, where the unknown sample is usually a molecular absorption spectrum, here the sample is the sample FC with its modes amplitudes/phases, while the LO-FC ones are known and taken as reference. As we will see in the following, the precision of the phase measurement is below $1 \degree$, achieved thanks to the combination of an active stabilization of the sample FC spacing with an electronic cancellation of common-mode noise, significantly improving the phase stability. As a consequence, we have been able to monitor the phase coherence at a time-scale of seconds, while typical DCS setups do not average the signal for longer than tens of ms \cite{Chen:2018}. 

In fig.~\ref{fig:dual-comb_setup} the setup used for dual-comb multi-heterodyne detection is sketched. According to eq.~\ref{eq:FCfield}, the frequency of a FC mode can be written as $f_n = n f_\text{s} + f_\text{o}$. In order to measure the Fourier phases, it is mandatory to get rid of $f_\text{s}$ and $f_\text{o}$ fluctuations, for both the involved FCs. $f_\text{s}$ can usually be detected as BN between the FC modes (IBN), and its fluctuations can be canceled by implementing a phase-locked loop (PLL) RF chain acting on the FC, having a stable clock as reference (see section \ref{sec:MIR_offset_subtr}). 

For FCs with a suitable second actuator, a referencing scheme implementing a second PLL chain, controlling $f_\text{o}$, is used \cite{Jones:2000}. In case the second actuator is missing or inefficient, $f_\text{o}$ noise can be removed from the RF-FC by electronically mixing a selected reference BN (RBN) with the other BNs (see fig.~\ref{fig:dual-comb_setup}d) \cite{Cappelli:2016}. The resulting RF difference BNs are free of common-mode frequency noise $\delta f_\text{o}$ (see fig.~\ref{fig:dual-comb_setup}d and f, and sections \ref{sec:MIR_offset_subtr} and \ref{sec:THz_offset_subtr}). In this regard, since common-mode noise provides no information on intrinsic FC operation, all the results presented here are common-mode-noise-free. 

In order to get simultaneous information about the Fourier phases of the modes, the RF-FC signal is acquired as time trace (two quadratures, I and Q) with a spectrum analyzer (see fig.~\ref{fig:dual-comb_setup}e) using the highest sampling rate available (75~MS/s), which results in a Nyquist frequency of $\pm 37.5$~MHz. The RF-FC time traces can be split into a variable number of consecutive time frames. This allows to observe the evolution of the phases at different time scales. At shorter time scales, i.e. 1~ms, the limiting factor is the BNs signal-to-noise ratio (S/N), whereas at longer time scales is the buffer capacity of the digitiser, the limit being about 2~s. Each frame is processed by means of an automated FFT analysis routine, retrieving frequency, amplitude and phase of all the BNs. This procedure is at the basis of the FACE method. The detailed procedure used for analyzing these signals is described in section \ref{sec:data_analysis_proc}.

\subsection{fs-pulsed FCs}

\begin{figure}[!htbp] 
\begin{center} 
\includegraphics[width=1.0\textwidth]{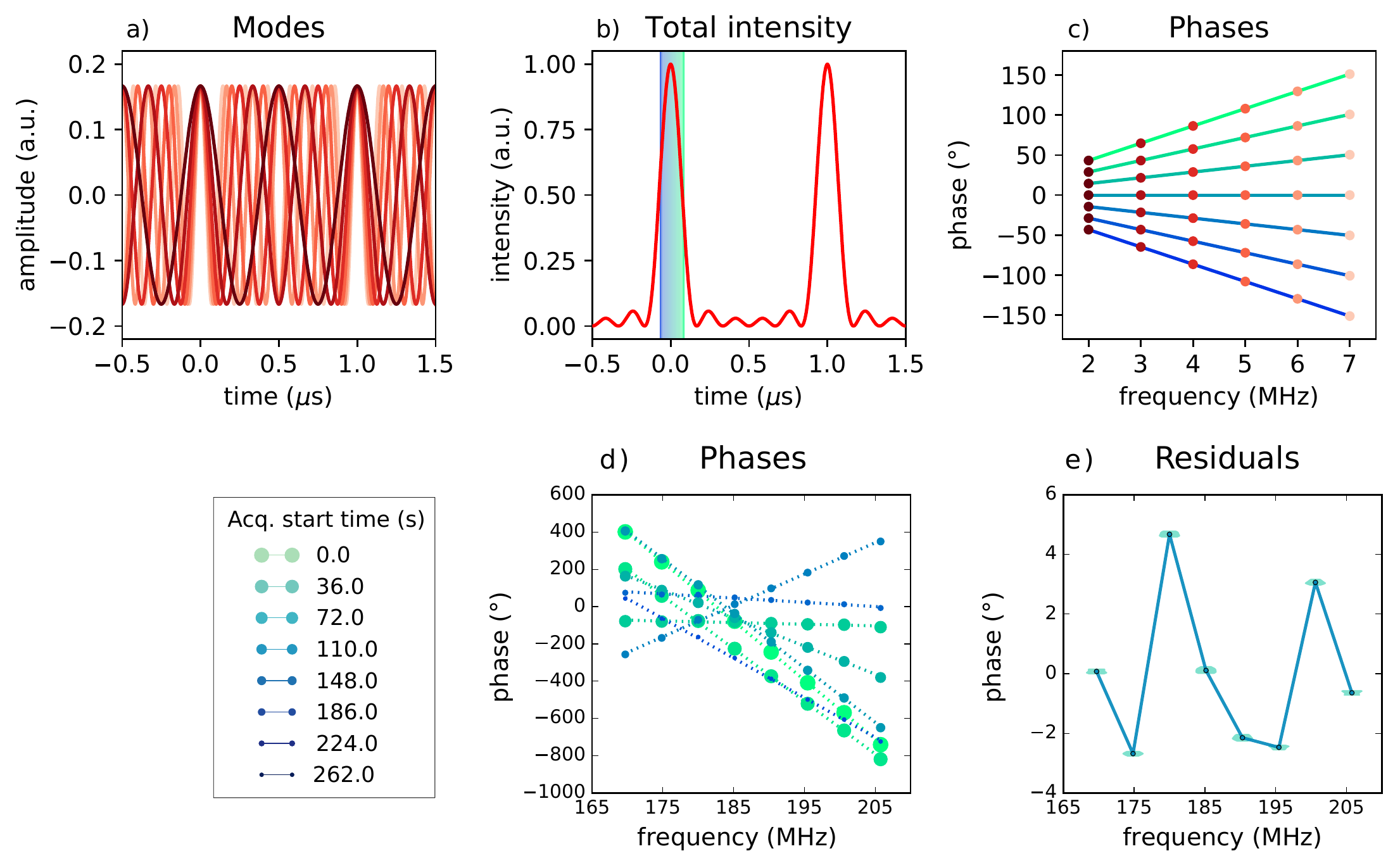}
\caption[]{fs-pulsed FC phase relation. \newline Top. Simulation of a pulsed FC emission. a)~Representation of the modes of a FC in the time domain. b)~Simulation of the intensity profile of a FC emission in the time domain (pulses). c)~Time evolution of the phases of the single modes around the pulse peak (see eq.~\ref{eq:FCfield}). The phases evolve linearly in time. In correspondence of the pulse top $\phi_{m} = 0$, otherwise they lie on a beeline, whose slope depends on the observation time. Time (and frequency) units are arbitrarily chosen. \newline 
Bottom. Phase analysis of the two commercial fs-pulsed FCs. d)~Measured BNs phases given by the FFT routine, just unwrapped. Beelines representing the linear fits are shown. The varying slope is due to the asynchronous acquisition. e)~Fit residuals of the measured BNs phases. The lines represent the phase relation. 
} 
\label{fig:phases_simul_NIR-combs} 
\end{center} 
\end{figure}

Fig.~\ref{fig:phases_simul_NIR-combs}-top shows a simulation of a pulsed FC emission, whose modes constructively interfere for pulse formation (fig.~\ref{fig:phases_simul_NIR-combs}a and b). Fig.~\ref{fig:phases_simul_NIR-combs}c shows the phases of the modes presented in fig.~\ref{fig:phases_simul_NIR-combs}a, color coded. The phase relation is linear, with an evolving slope depending only on the observation time. The first experiment here presented, involving two commercial fs-pulsed FCs, aims to probe the precision of our technique on this well-defined phase pattern.

The experimental details regarding the pulsed FCs can be found in section \ref{sec:pulsed_meth}, while in fig.~\ref{fig:phases_simul_NIR-combs}d the phases retrieved for 8 consecutive 1-s-long acquisitions are plotted, with a dead time of 38 s between consecutive acquisitions, thus covering an overall 4-minutes time interval. The peak centered at 185~MHz is the RBN for $f_\text{o}$ subtraction. The relevant time-independent phase information (Fourier phases) is provided by the fit residuals (see fig.~\ref{fig:phases_simul_NIR-combs}e) representing the phase relation. Computing the standard deviation of each BN phase it is possible to estimate its time stability over the considered time interval. The phase of each BN is stable within $0.20 \degree$, basically depending on the S/N of the BN. 

The results show that the phase relation among the observed modes of the two classical FCs is not completely flat, the phases falling within a range of $8.0 \degree$ (see fig.~\ref{fig:phases_simul_NIR-combs}e). In this regard, in section \ref{sec:phases_simulation} a simulation of a pulsed FC emission, obtained by varying the scattering of the randomly-generated phase values assigned to the modes, is shown. Pulses with a good contrast can still be generated with a scattering of the phases of $45 \degree$, while scattering values exceeding $180 \degree$ severely degrade the pulsed regime.

\subsection{Mid-IR and THz QCL-combs} 

The specific, almost-flat, phase relation of pulsed FCs is particularly handy when using these sources as LOs for measuring the unknown phase relation of a sample FC, e.g. mid- and far-IR QCL-combs. For measurements in the mid-IR and THz regions, convenient pulsed LO-FCs can be obtained by difference frequency generation (DFG-FCs). In the measurements on QCL-combs presented in this paper, we have used the FCs described in section \ref{sec:pulsed_meth} as pumping sources for down-conversion setups to mid- and far-IR \cite{Galli:2013d,Galli:2014a,Campo:2017,Consolino:2012,Bartalini:2014}. Due to the generation process, DFG-FCs inherit the same $f_\text{s}$ as the pumping FCs. A complete QCL-combs descriptions can be found in section \ref{sec:qcls_desc}. QCL-combs $f_\text{s}$ can be electrically extracted from the laser chip. This signal is used in a PLL for actively phase locking $f_\text{s,QCL}$ to a RF oscillator (referenced to the quartz/Rb/GPS clock) by modulating the QCL bias current. Such phase-locking removes the frequency noise contribution on the detected BNs coming from $f_\text{s,QCL}$. Again, a single RBN is isolated and mixed with all the other BNs in order to remove common-mode noise ($\delta f_\text{o}$). All the RF chains are described in detail in sections \ref{sec:MIR_offset_subtr} and \ref{sec:THz_offset_subtr}. 
\begin{figure}[!htbp]
\begin{center}
\includegraphics[width=0.9\textwidth]{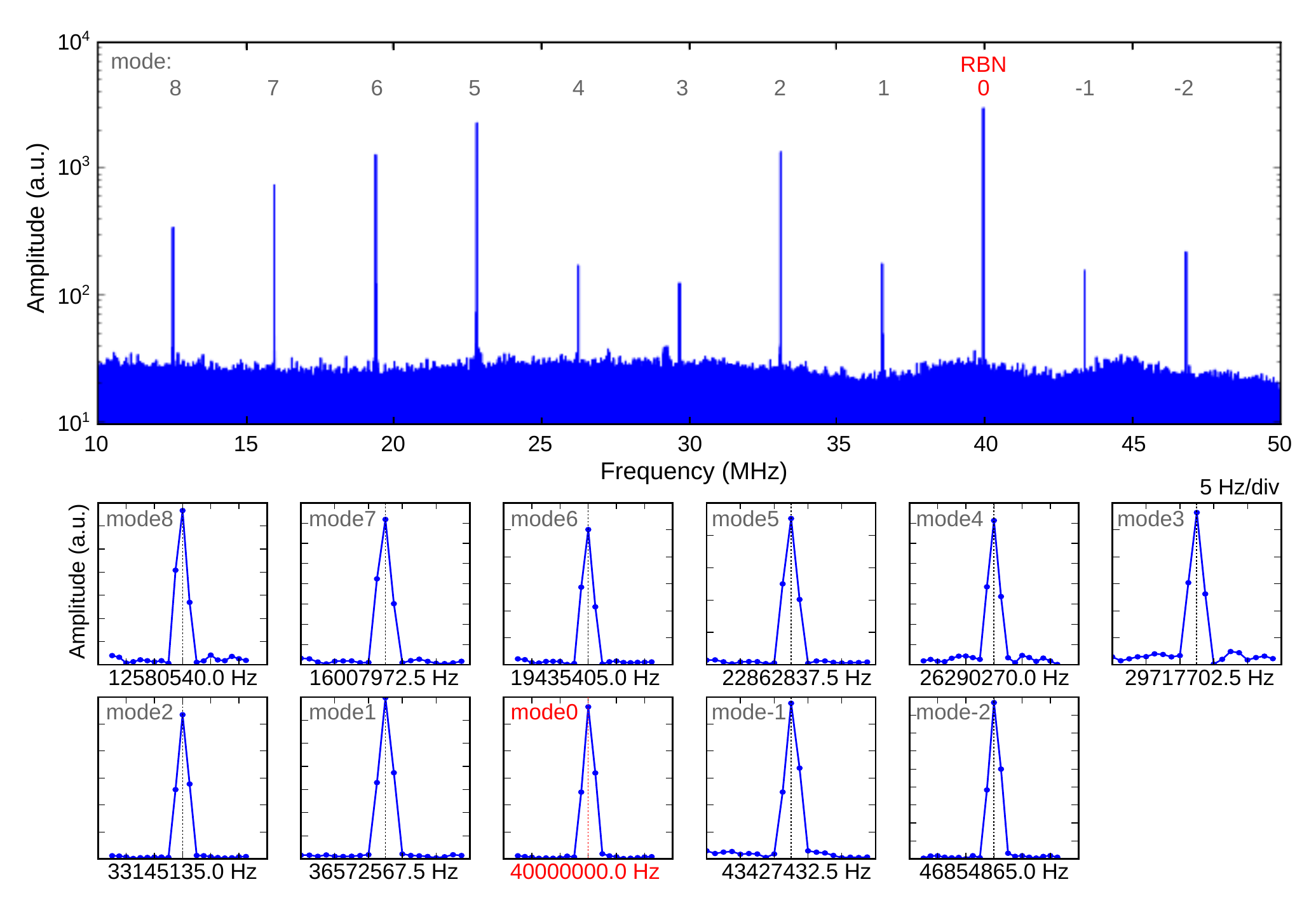}
\caption[]{FFT amplitude spectrum of the RF-FC related to the THz QCL-comb. \newline Top. Analysis performed on a 800~ms-long frame. The 11 BNs are clearly visible, each generated by the beating between one mode of the QCL-comb and one mode of the THz LO-FC. The BN labelled with 0 is the RBN. \newline 
Bottom. The stability of the BNs frequency is evidenced by their 2.5~Hz full width at half maximum, limited by the acquisition time. } 
\label{fig:THzQCL_dual-comb_spectrum}
\end{center}
\end{figure}
The effect of the PLL and of the common-mode noise removal is reported in fig.~\ref{fig:THzQCL_dual-comb_spectrum}, where an acquisition of the RF-FC amplitude spectrum related to the THz QCL-comb is shown. The spectrum is made of a series of equally-spaced narrow peaks, confirming FC operation. For this reason, the technique can also detect in real-time a non-perfect FC behaviour, or transitions in or out of FC operation. Zooming over each BN (fig.~\ref{fig:THzQCL_dual-comb_spectrum}-bottom) reveals that the BNs width is always limited by the analysis resolution bandwidth (RBW -- 1.25~Hz in this case). This result is the first direct evidence of the remarkable stability, at 1-Hz level, of the BNs frequencies.


For the mid-IR QCL-comb, whose $f_\text{s,QCL}$ is about 7.060~GHz, in order to optimize the multi-heterodyne dual-comb BN detection, $f_\text{s-LO}$ is set to 1.00780~GHz, leading to $f_\text{s,BNs} = 5.467600$~MHz with $k = 7$ (eq.~\ref{eq:RFspacing}), enabling the simultaneous acquisition of 12~BNs within an actual frequency span of 61~MHz. 
\begin{figure}[!htbp]
\begin{center}
\includegraphics[width=1.0\textwidth]{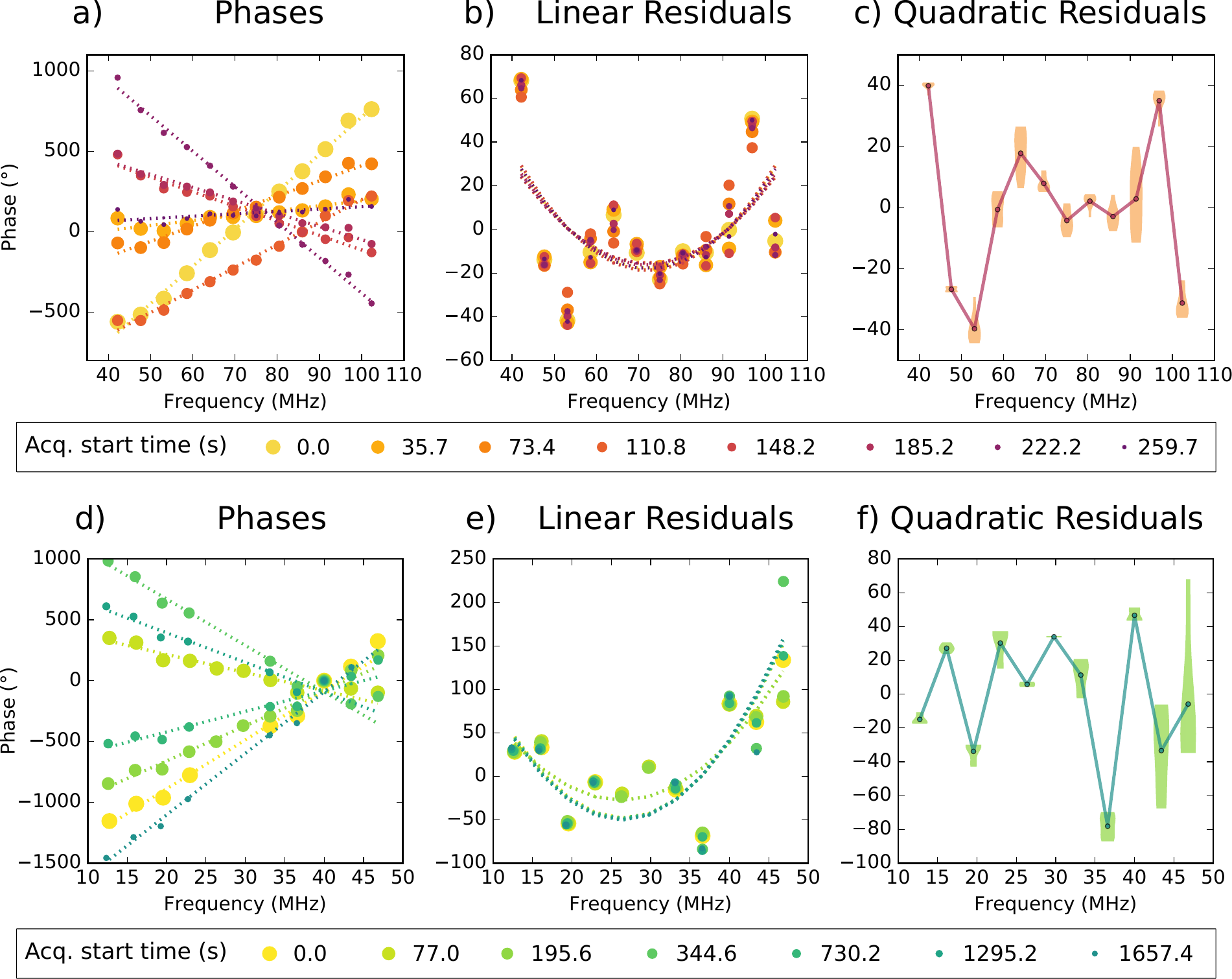}
\caption[]{Phases of the BNs related to the QCL-combs computed by the FFT routine. \newline 
Top. Mid-IR QCL-comb phases. a)~BNs phases given by the FFT routine, just unwrapped. Beelines representing the linear fits are shown. b)~Linear fit residuals representing the Fourier phases. Parabolas representing the quadratic fits are shown. c)~Averaged quadratic fit residuals. The lines represent the phase relation. The error bar thickness represents the single phase distribution (violin plot). \newline 
Bottom. THz QCL-comb phases. d)~BNs phases given by the FFT routine. Beelines representing the linear fits are shown. e)~Linear fit residuals representing the Fourier phases. Parabolas representing the quadratic fits are shown. f)~Averaged quadratic fit residuals. The lines represent the phase relation. The error bar thickness represents the single phase distribution (violin plot). }
\label{fig:QCLph-all-acqs}
\end{center}
\end{figure}
In fig.~\ref{fig:QCLph-all-acqs}-top the phases retrieved for 8 consecutive 1s-long acquisitions covering a total of 4 minutes are compared. Here, the linear fit residuals (fig.~\ref{fig:QCLph-all-acqs}b) show an evident parabolic trend with a concavity of $(0.048 \pm 0.005) \degree/\text{MHz}^2$, corresponding to a group delay dispersion (GDD) of $ (12.9 \pm 1.3)~\text{ps}^2/\text{rad}$. The quadratic fit residuals (fig.~\ref{fig:QCLph-all-acqs}c) clearly show a fixed phase relation. The obtained phases are stable within $8.0 \degree$. The only exception is the phase of the BN at 91.4~MHz, that is clearly less stable than the others, with fluctuations of about $14 \degree$.


For the THz QCL-comb $f_\text{s,QCL} \approx 19.784$~GHz, while $f_\text{s-LO}$ is set to 250.39~MHz, yielding $f_\text{s,BNs} = 3.4274325$~MHz with $k = 79$ (eq.~\ref{eq:RFspacing}), enabling the simultaneous acquisition of 11~BNs, corresponding to all the modes emitted by the device. Fig.~\ref{fig:QCLph-all-acqs}-bottom shows the phases retrieved in 7 different 1s-long acquisitions taken in a time interval of about 30 minutes. In this case, the linear fit residuals (fig.~\ref{fig:QCLph-all-acqs}e) show a parabolic trend with a concavity of $(0.45 \pm 0.09) \degree/\text{MHz}^2$, yielding a GDD of $(6.0 \pm 1.2)~\text{ps}^2/\text{rad}$, while the quadratic fit residuals of each mode (fig.~\ref{fig:QCLph-all-acqs}f) are stable within about $10 \degree$, excluding the last two modes that are clearly less stable.

\section{Discussion} 

Considering the stability of the measured Fourier phases, the newly-proposed combination of active stabilization and electronic subtraction of common-mode noise allowed to achieve a precision of $0.2 \degree$ at minutes timescales. As a consequence, this is the level of coherence that has been proven for the fully-stabilized commercial fs-pulsed FCs. 

On the other hand, when studying QCL-combs, our approach works also with the active stabilization of $f_\text{s}$ only. This minimizes the interaction with the QCL-comb, and allows its study in a condition as close as possible to unperturbed operation. In this case, the observed standard deviations are more than one order of magnitude larger than the ones in pulsed FCs, in the order of $8 - 10 \degree$ at tens-of-minutes timescale (see table \ref{tab:res_summ}). 

\begin{table}
\begin{center}
\vspace{\baselineskip}
\begin{small}
\begin{tabular}{| A | B | B |}
\hline 
 & mid-IR QCL-comb & THz QCL-comb \\ \hline
single phases $\sigma~( \degree)$ & 8.0 & 10.0 \\ \hline
phases scattering $( \degree)$~$^\triangle$ & 120 & 300 \\ \hline
phases concavity $( \degree/\text{MHz}^2)$ & $0.048(5)$ & $0.45(9)$ \\ \hline
GDD~$^*$ $( \text{ps}^2/\text{rad} )$ & $12.9(1.3)$ & $6.0(1.2)$ \\ \hline
\end{tabular}
\end{small}
\end{center}
\caption[]{Summary of the parameters related to the QCL-combs phases. \newline As a benchmark, for the near-IR FCs the standard deviation on the single phases is $0.20 \degree$, while the phases scattering is $8.0 \degree$. \\ 
$^*$~GDD is computed from the phases concavity as $\text{GDD} = [C_\text{$\phi$}/(180 \cdot 4 \pi)] \cdot (f_\text{s,BNs}/f_\text{s,sample})^2$. \\ 
$^\triangle$ On linear residuals. }
\label{tab:res_summ}
\end{table}

Concerning the phase relation, noticeably, the scattering of THz QCL-comb phases (linear residuals) encompasses a range of $300 \degree$, a value which cannot be associated to short-pulses operation (see section \ref{sec:phases_simulation}). On the other hand, the scattering of the mid-IR QCL-comb phases covers a range of $120 \degree$, a value that would still allow generation of short pulses with a good contrast. However, only a limited number of mid-IR FC modes could be observed in this case ($\approx 1/10$ of the total number), possibly leading to an underestimation of the overall phases scattering. 

The quadratic term found in the phase relations of the two QCL-combs (see fig.~\ref{fig:QCLph-all-acqs}b and e, and table \ref{tab:res_summ}) can be attributed to chirping in either DFG-FCs pulses or QCL-combs emission. However, the obtained GDD values are very similar in the two cases, while the two experimental setups are completely different. Moreover, the pump fs lasers used for DFG-FCs generation are optimized for the shortest possible pulse duration, i.e. the smallest chirp in the emission. As a consequence, it is highly probable that the measured chirp is due to the QCL-combs. A comparison of the obtained GDD values with what reported in ref.~\cite{Singleton:2018} for a dispersion-compensated mid-IR QCL-comb suggests that the devices described in this work operate in a non-pure linear-chirp regime. In fact, the measured GDD values for these two devices are smaller (2 and 6 times for mid-IR and THz QCL-combs, respectively) compared to the theoretical values calculated following the approach therein presented for maximally-chirped pulse operation. This evidence suggests that dispersion compensation significantly influences the operation of QCL-combs, favoring maximally-chirped operation, and that the devices described in this work operate in a different regime, as confirmed by the following analysis. 

%
\begin{figure}[!htbp]
\begin{center}
\includegraphics[width=1.0\textwidth]{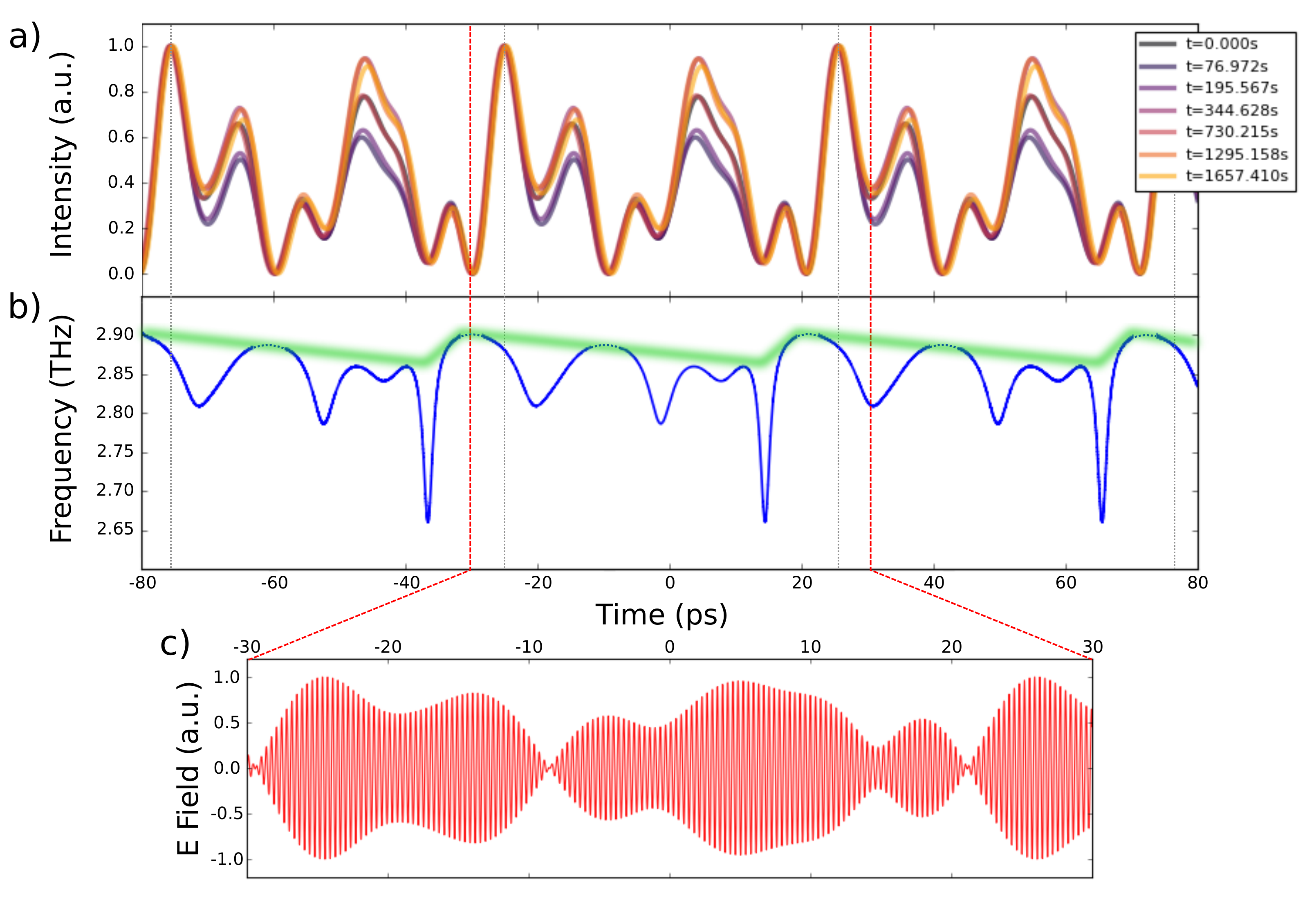}
\caption[]{THz QCL-comb emission. \newline a) Reconstruction of the THz QCL-comb output intensity, based on the 7 sets of measurements of fig.~\ref{fig:QCLph-all-acqs}-bottom. b) Calculated QCL-comb instantaneous frequency from Hilbert transform of the electric field. The dotted lines represent the areas of higher uncertainty, due to low emitted intensity, while the shaded green trace identifies a linear chirp component in the device emission. c) Detail of the QCL-comb emitted electric field. }
\label{fig:field_reconstr}
\end{center}
\end{figure}

For the THz QCL-comb we are able to simultaneously measure the complete set of emitted modes. Starting from the measured frequencies, amplitudes and phases of the modes, we are able to reconstruct the electric field, the intensity profile and the instantaneous frequency of the QCL-comb emission, as shown in fig.~\ref{fig:field_reconstr}. The intensity profiles are obtained from seven reconstructions, based on the independent measurements shown in fig.~\ref{fig:QCLph-all-acqs}-bottom. The emission profiles are roughly the same from measurement to measurement, as already suggested by the consistency of the phases over the entire 30 minutes-long observation period. Noticeably, the QCL-comb emission deviates from a pulsed emission, even though its amplitude is deeply modulated during the round trip time. The frequency of the QCL-comb is calculated by Hilbert transform of the electric field and confirms that the THz QCL-comb operates in a hybrid amplitude-/frequency-modulated regime \cite{Benea:2017}. In fact, in the region around 2--10~ps (see fig.~\ref{fig:field_reconstr}), a pulse-like shape coherently corresponds to an almost constant frequency (amplitude modulation). In the remaining time intervals, an important frequency modulation clearly emerges. This sort of behaviour is found also in similar devices, both in THz QCL-combs, with opposite operation modes observed for two distinct emission lobes \cite{Burghoff:2015}, and in dispersion-compensated mid-IR QCL-combs \cite{Singleton:2018}, where the strong linear chirp is absent in correspondence of the pulse-like emission. In addition, for the device under investigation the emission intensity and the instantaneous frequency appear to be related. This might be due to phenomena induced by the high third-order nonlinear coefficient originating the four-wave mixing, such as self-phase modulation.  At the same time, the retrieved electric field, shown in fig.~\ref{fig:field_reconstr}c, is similar to that found with a completely different technique for actively mode-locked devices \cite{Wang:2015}. 

In conclusion, we have introduced the FACE method for real-time monitoring of the Fourier phases of a generic FC by direct comparison with those of a metrological-grade FC, using a phase-stable dual-comb multi-heterodyne detection scheme and the Fourier analysis. This is particularly interesting for new-generation FCs, where the phase relation among the modes can be non-trivial. The knowledge of the phase relation provides the basis for future implementation of programmable pulse shaping \cite{Ferdous:2011}. Moreover, the analysis covers time scales ranging from ms to tens of minutes, and can be used to validate theoretical models describing FC formation mechanisms \cite{Khurgin:2014,Villares:2015a,Faist:2016,Tzenov:2016,Del'Haye:2014,Herr:2013}. The remarkable phase stability attained for the QCL-combs modes conclusively proves the high coherence characterizing their emission, paving the way to significant improvements for present applications, e.g. broadband phase-sensitive spectroscopy and metrological FC signals distribution, but also to unpredictably new setups and measurements.

\renewcommand{\thesection}{M}

\section{Methods}
\label{sec:methods}

\subsection{IQT acquisition and phases retrieval}
\label{sec:data_analysis_proc}

A real-time FFT approach is required in order to obtain a simultaneous information about the Fourier phases of the modes (FACE method). For this reason the RF-FC signal is acquired as time trace (in its two quadratures I and Q, related to a reference RF LO) with a spectrum analyzer using the highest sampling rate available (75~MS/s). A dedicated FFT routine computes the Fourier transform giving the amplitude and the phase spectrum of the signal. Successively, a specifically-developed routine identifies the FFT frequencies of the BNs in the amplitude spectrum and collects the related phases from the phase spectrum. For the FFT algorithm the resolution bandwidth is given by $\text{RBW} = 1/\tau_\text{L}$, where $\tau_\text{L}$ is the length of the time trace.

\subsection{Pulsed fs FCs}
\label{sec:pulsed_meth}

The dual-comb multi-heterodyne detection setup for measuring the phases of the modes of the two commercial FCs is here described (see fig.~\ref{fig:dual-comb_setup}a in the main text). The sample FC is a visible/near-IR FC generated by a Ti:Sa laser centered at 810~nm, operating in pulsed passive mode-locking, with $f_\text{s,sample}$ tunable around 1~GHz, spectrally broadened to more than one octave (500--1100~nm) by a photonic-crystal fiber. The LO FC is a near-IR FC generated by an Er-doped fiber laser centered at 1550~nm, operating in pulsed passive mode-locking, with $f_\text{s,LO} = 250$~MHz, and broadened to achieve a 1000--2000~nm spectral coverage by a highly-nonlinear fiber. The optimal spectral superposition between the two FCs falls in the wavelength region around 1064~nm. There, the two FCs are optically filtered by means of a high-resolution 2-m-focal-length Fastie-Ebert monochromator SOPRA \cite{Mazzacurati:1988}, opportunely modified for operating in the near-IR with an echelle grating (720~g/mm) working at first order. Optical filtering ensures the detection of only the BNs of interest, contemporarily increasing their signal to noise ratio (S/N). 
$f_\text{s,sample}$ and $f_\text{s,LO}$ are phase stabilized and tuned in order to have $f_\text{s,BNs} = 5.152980$~MHz with $k = 4$ (see eq.~\ref{eq:RFspacing}). In this configuration, a RF-FC is detected by using a fast photodetector, one RBN is isolated with a narrow-band RF filter and $f_\text{o}$ is canceled out by RF mixing (see section \ref{sec:MIR_offset_subtr}). With $f_\text{s,BNs} = 5.152980$~MHz within a frequency span of 40~MHz 8~BNs can be simultaneously acquired. 

\section*{Acknowledgements}

The authors acknowledge financial support by: 
\begin{itemize}
 \item Ministero dell’Istruzione, dell’Università e della Ricerca (Project PRIN-2015KEZNYM ``NEMO -- Nonlinear dynamics of optical frequency combs''); 
 \item European Union’s Horizon 2020 research and innovation programme (Laserlab-Europe Project [grant No 654148], CHIC Project [ERC grant No 724344], ULTRAQCL Project [FET Open grant No 665158] ``Ultrashort Pulse Generation from Terahertz Quantum Cascade Lasers'', and Qombs Project [FET Flagship on Quantum Technologies grant No 820419] ``Quantum simulation and entanglement engineering in quantum cascade laser frequency combs''); 
 \item European ESFRI Roadmap (``Extreme Light Infrastructure'' -- ELI Project);
 \item Swiss National Science Foundation (SNF200020-165639). 
\end{itemize}

\section*{Author contributions}

F.C. and S.B. conceived the experiment. L.C., F.C., G.C., I.G., M.S.d.C., A.C., P.C.P. and R.E. performed the measurements. F.C., G.C., R.E. and S.B. analyzed the data. L.C. and F.C. wrote the manuscript. G.C., D.M., M.S.d.C., P.C.P., R.E., S.B., G.S., J.F. and P.D.N. contributed to manuscript revision. J.F., G.S., M.R. and M.B. provided the quantum cascade lasers. L.C., F.C., D.M., P.C.P., R.E., S.B., G.S., J.F. and P.D.N. discussed the results. L.C. and F.C. contributed equally to this work. All work was performed under the joint supervision of P.D.N. and S.B..

\section*{Competing interests}

The authors declare no competing financial interests.

\section*{Data availability}

The data that support the plots within this paper and other findings of this study are available from the corresponding authors upon reasonable request.


\clearpage

\renewcommand{\thesection}{S}
\renewcommand\thefigure{\thesection\arabic{figure}}    
\setcounter{figure}{0}

\section{Supplementary information}
\label{sec:supplementary}

\subsection{Quantum cascade laser devices}
\label{sec:qcls_desc}

The mid-IR QCL investigated in this work is a broad-gain Fabry-P\'erot device operating at a temperature $T = 18 \degree$C, with a threshold current $i_\text{th} = 600$~mA, emitting at a wavelength $\lambda = 4.7~\mu$m. Within the 645--760-mA driving current range the QCL generates a FC \cite{Cappelli:2016} characterized by about 100 modes with $f_\text{s,QCL} = 7.060$~GHz and an overall emitted power of 40~mW per facet. 
\begin{figure}[!htbp]
\begin{center}
\includegraphics[width=0.7\textwidth]{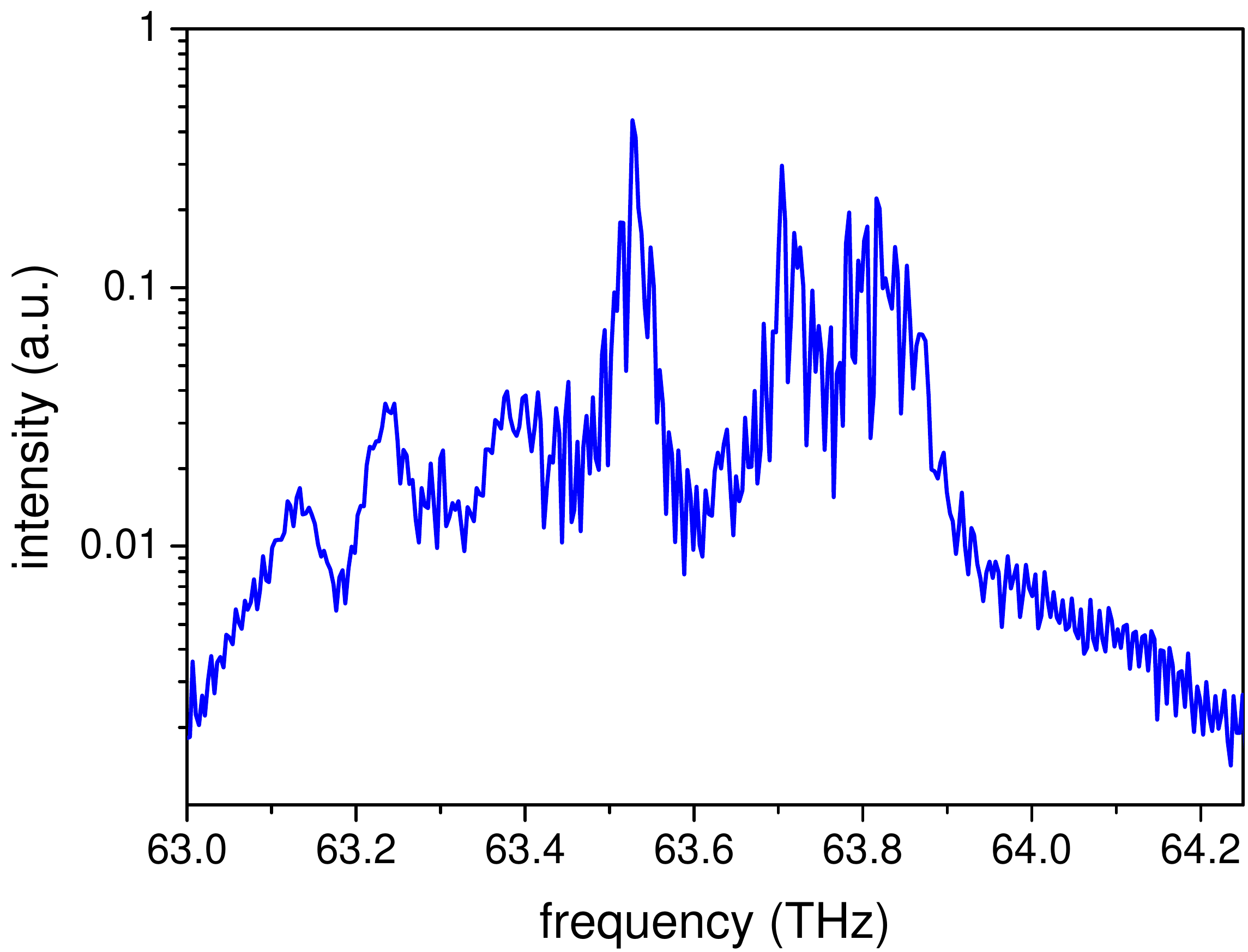}
\caption[]{Mid-IR QCL-comb. \newline Amplitude spectrum acquired with an optical spectrum analyzer. }
\label{fig:QCL654mAspectrum}
\end{center}
\end{figure}
In fig.~\ref{fig:QCL654mAspectrum} the QCL-comb spectrum acquired with an optical spectrum analyzer is shown. 

\begin{figure}[!htbp]
\begin{center}
\includegraphics[width=1.0\textwidth]{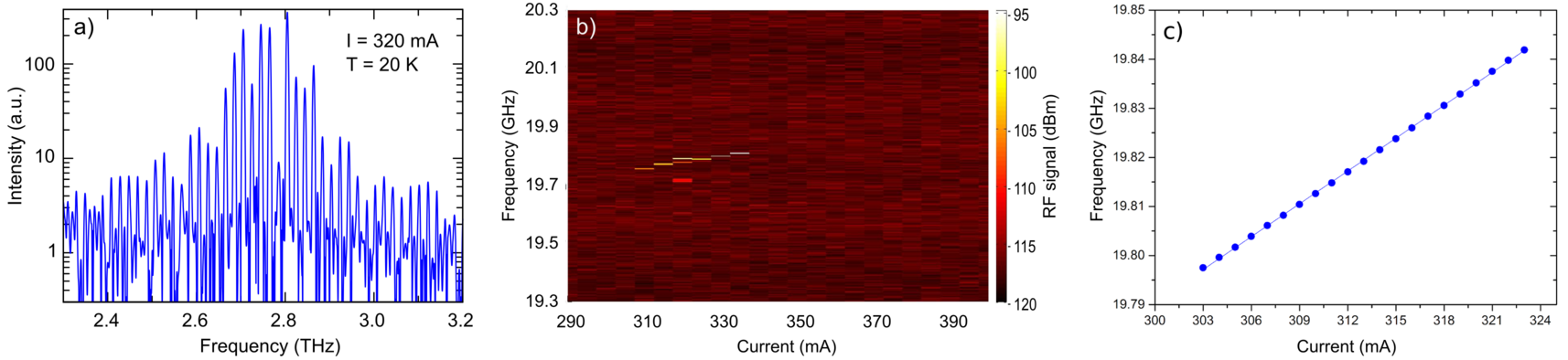}
\caption[]{THz QCL-comb. \newline a) Amplitude spectrum at the operating current of 320 mA and at a temperature T = 20~K. b) Intermodal beat note as a function of the injected current electrically detected with a bias-tee on the QCL bias line. c) Detail of the tunability of the IBN, with a slope of 2.23~MHz/mA. }
\label{fig:THzQCLspectrum}
\end{center}
\end{figure}

The THz QCL-comb is a heterogeneous cascade laser device emitting in the 2.5--3.1~THz spectral region (see fig.~\ref{fig:THzQCLspectrum}a). It is based on a double-metal ridge waveguide, defined via dry etching, of length 2~mm and width $60~\mu$m, with lateral side absorbers for enhanced transverse mode suppression \cite{Bachmann:2016}. When operated in continuous wave (CW) at $T = 20$~K, it displays a clear and narrow intermodal beat note (IBN) around 19.8~GHz in the current range 305--335~mA, as shown in fig.~\ref{fig:THzQCLspectrum}b. In this regime, the IBN width is in the order of tens of kHz, and has a tunability of 2.23~MHz/mA (see fig.~\ref{fig:THzQCLspectrum}c). It is worth noting that, in the range of current where the IBN signal is detected (305--335~mA), the spectrum of the QCL shows 11 intense modes, while the other emitting modes have an intensity more than 10 times lower.

\subsection{Mid-IR: $f_\text{s}$ phase locking and $f_\text{o}$ fluctuations suppression}
\label{sec:MIR_offset_subtr}

\begin{figure}[!htbp] 
\begin{center}
\includegraphics[width=0.8\textwidth]{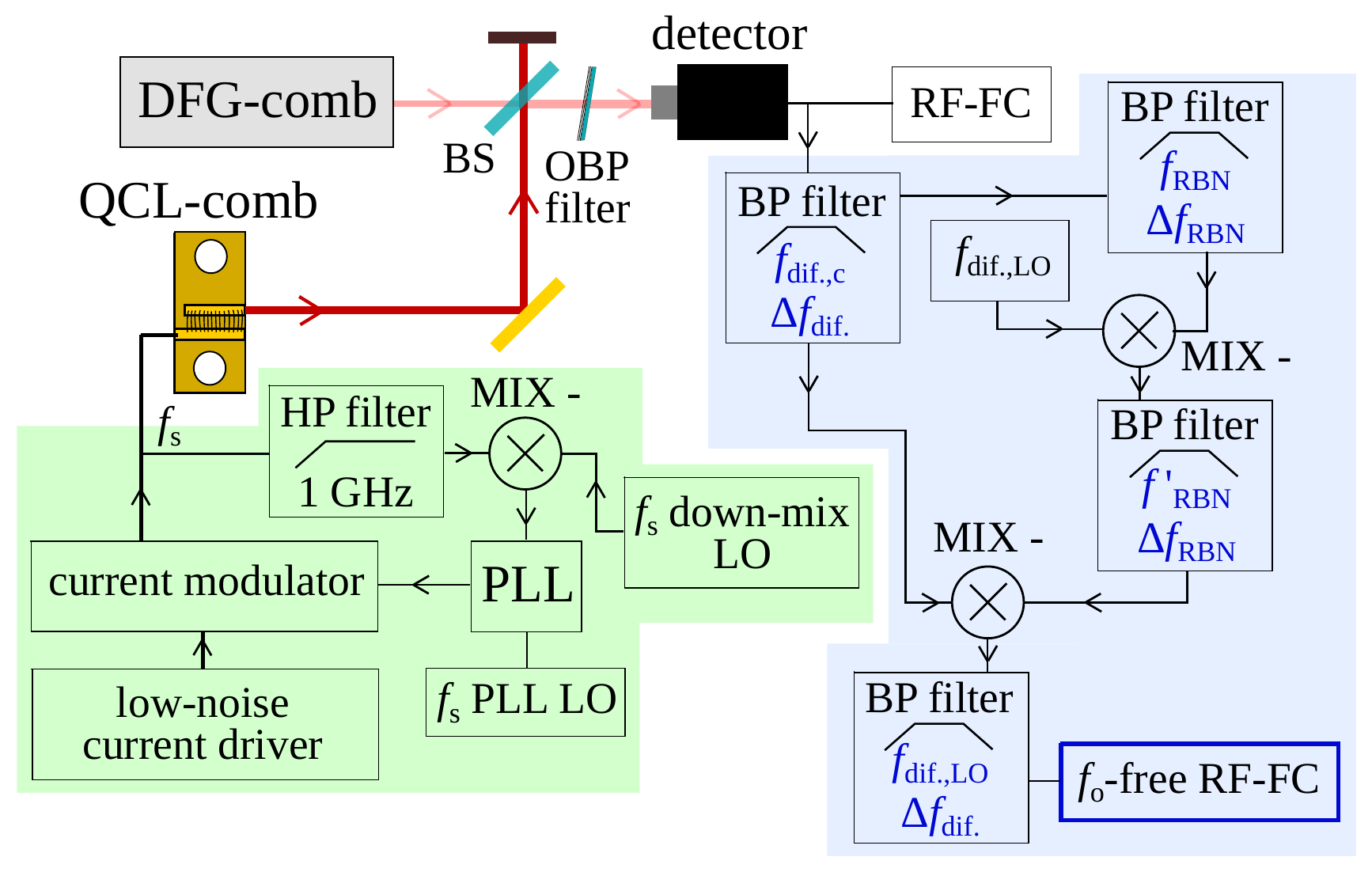}
\caption[]{Mid-IR QCL-comb. \newline Electronic setup for $f_\text{s}$ stabilization and $f_\text{o}$ cancellation. Left (green background): RF chain for $f_\text{s,sample}$ phase locking. Right (blue background): RF chain for common-mode noise ($f_\text{o}$) suppression. BS: beam splitter. OBP: optical band-pass filter. BP: RF band-pass filter. For band-pass filters both the center frequency and the bandwidth are indicated. HP: RF high-pass filter. MIX -: RF mixer (providing the difference frequency). }
\label{fig:dual-comb_RFmix_setup}
\end{center}
\end{figure}

\begin{figure}[!htbp]
\begin{center}
\includegraphics[width=0.9\textwidth]{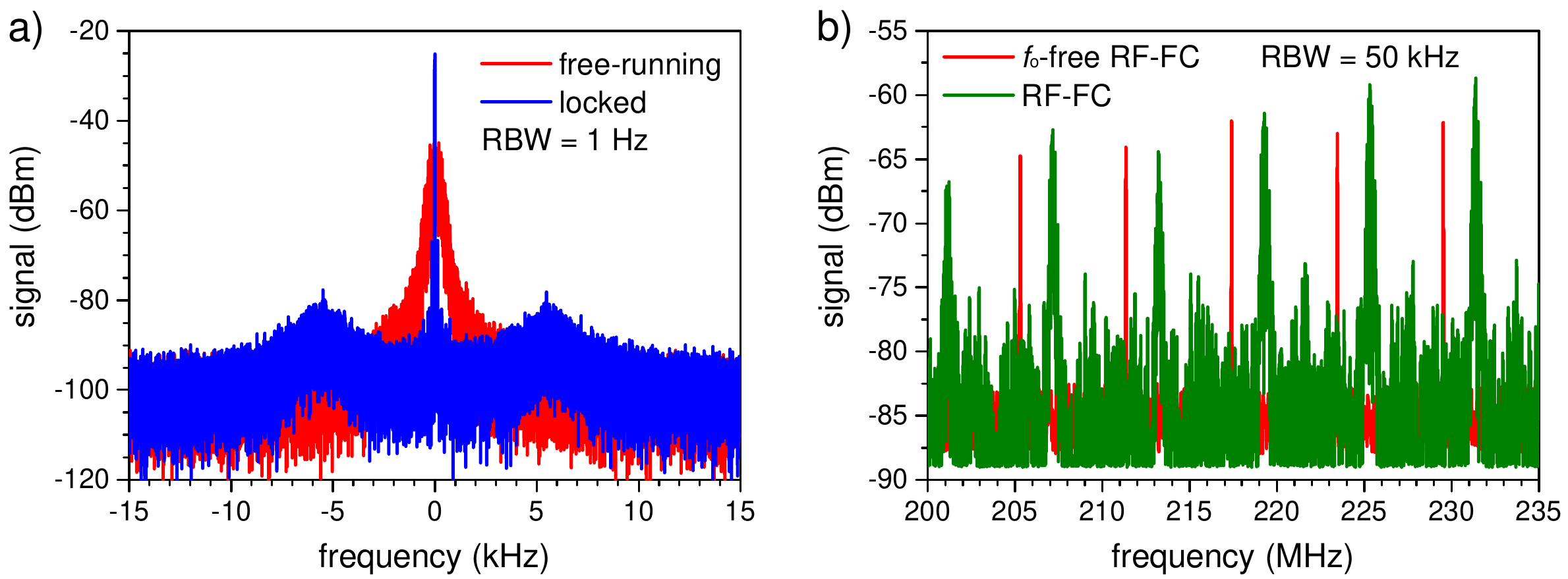}
\caption[]{Mid-IR QCL-comb. \newline a) $f_\text{s}$ signal in free-running and locking operation. The locking frequency is 7.06006750~GHz, the locking efficiency is 99\% and the locking bandwidth is 5~kHz. b) Comparison between the original RF-FC and the $f_\text{o}$-free RF-FC. For the latter the BNs are resolution-bandwidth limited. }
\label{fig:QCLfsfosub}
\end{center}
\end{figure}

An effective measurement of the Fourier phases of FC modes requires fully-stabilized signals, the residual frequency noise must be negligible within the adopted resolution bandwidth. For this purpose, a dedicated electronic setup for $f_\text{s}$ stabilization and $f_\text{o}$ cancellation has been realized (fig.~\ref{fig:dual-comb_RFmix_setup}). The setup is composed of two RF chains.  
The first one (bottom left -- green background) locks $f_\text{s,sample}$. When the sample FC is the QCL-comb, $f_\text{s,sample}$ (7.060~GHz) can be detected as RF signal directly contacting the laser chip. $f_\text{s,sample}$ is firstly filtered and down-mixed. At this point the attained signal can be phase locked to a RF LO (PLL LO) by using a PLL electronics for generating the error signal that is used to modulate the laser bias current. The current modulator consists of a control circuitry placed in parallel to the QCL. It is based on a field-effect transistor (FET) \cite{Cappelli:2016}, where the processed error signal is fed to the gate of the FET itself, that acts by proportionally decreasing the current flowing through the laser. The locking shows satisfactory stability and efficiency (99.0\%) with a bandwidth of 5~kHz (see fig.~\ref{fig:QCLfsfosub}a). 

The second RF chain (right side -- blue background) is used for removing the common-mode noise ($f_\text{o}$ fluctuations) from the dual-comb spectrum (RF-FC). The RF-FC spectrum is made of several BNs which are preliminarily filtered around $f_\text{dif.,c1}$ within a bandwidth $\Delta f_\text{dif.}$ slightly larger than the Fourier transform span. A single BN is selected as reference (RBN) by tightly filtering at $f_\text{RBN}$, then it is down shifted by RF mixing of a frequency $f_\text{dif.,LO}$ (75~MHz for the mid-IR QCL-comb) and again filtered around $f'_\text{RBN}$. The down-shifted RBN is then subtracted from all the other BNs by RF mixing. The result is a RF-FC centered at $f_\text{dif.,LO}$, with a bandwidth corresponding to $\Delta f_\text{dif.}$, where all the BNs are common-mode-frequency-noise free ($f_\text{o}$-free RF-FC). Thanks to the two RF chains, resolution-bandwidth-limited BNs with S/N~$\approx 10 - 30$ are attained (see fig.~\ref{fig:QCLfsfosub}b for a comparison between the original RF-FC and the $f_\text{o}$-free RF-FC). 

All the RF signals used as LOs are referenced to the primary frequency standard through a GPS-disciplined Rb-locked quartz oscillator. Along the RF chains, dedicated RF amplifiers keep the signals level at an adequate value for transmission and circuitry operation.

\subsection{THz: $f_\text{s}$ phase locking and $f_\text{o}$ fluctuations suppression}
\label{sec:THz_offset_subtr}

\begin{figure}[!htbp]
\begin{center}
\includegraphics[width=0.9\textwidth]{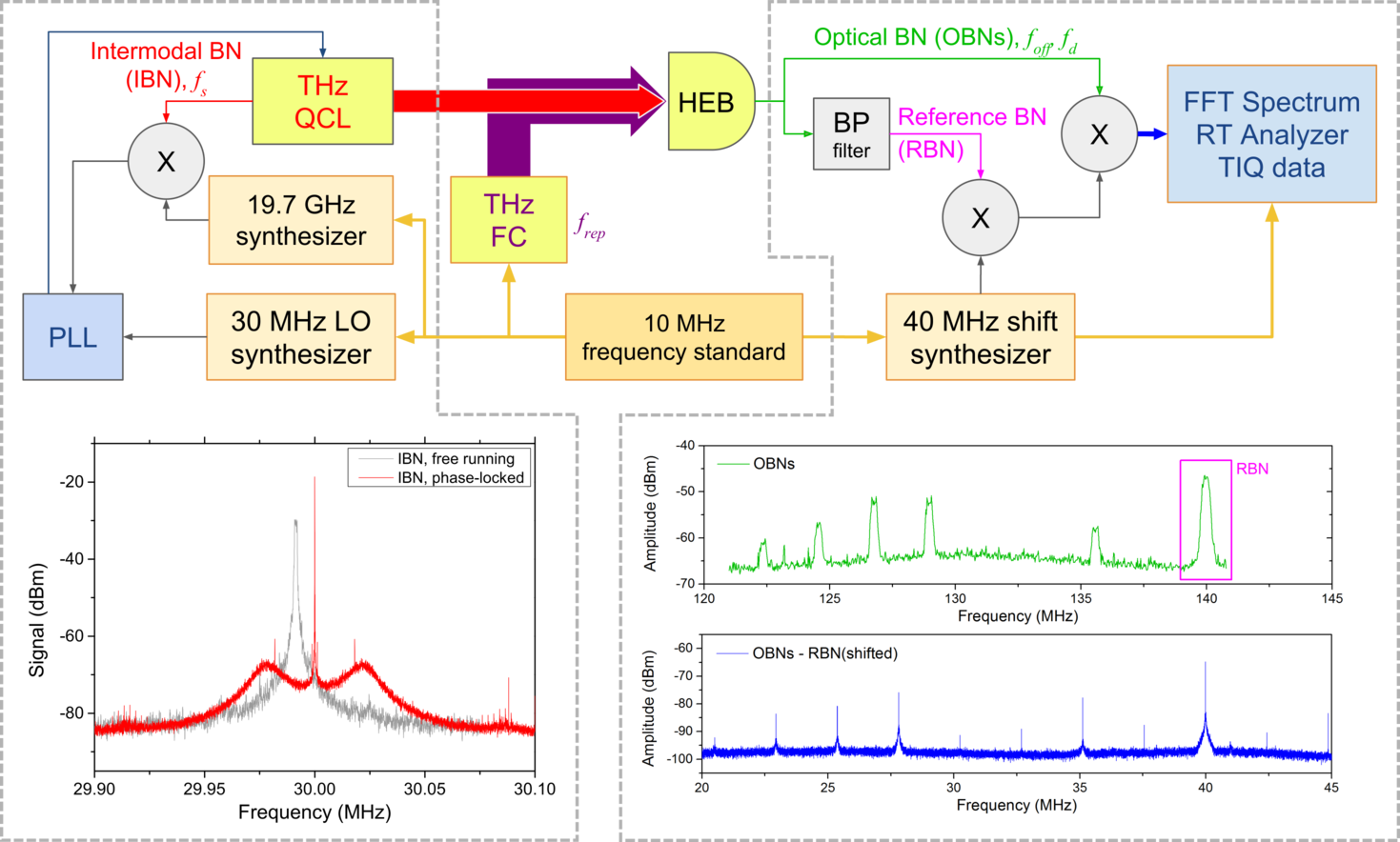}
\caption[]{THz QCL-comb. \newline Schematic setup for the hybrid active-passive stabilization of the dual-comb BNs. Left: An active PLL is closed onto the QCL bias current in order to actively stabilize the IBN, that is electrically extracted from the laser chip by means of a RF bias-tee. Right: A passive cancellation of the offset fluctuations is achieved by selecting one beat note (the RBN) and subtracting it (and therefore its frequency noise) from the others. }
\label{fig:THzQCLlock}
\end{center}
\end{figure}

For the stabilization of the THz QCL-comb, a similar setup has been used (fig.~\ref{fig:THzQCLlock}). Once again, it consists of two different sections, devoted to the suppression of spacing and offset jitters, respectively. 

For what concerns the stabilization of the modes spacing, an active phase-locking of the QCL IBN has been implemented (fig.~\ref{fig:THzQCLlock}-left). The IBN is electrically extracted from the QCL by mounting a bias-tee (Marki Microwave, mod. BT-0024SMG) inside the liquid helium cryostat, as close as possible to the device. The DC line feeds the current to the QCL, while the AC line is used to extract the IBN, that is acquired by a spectrum analyzer. The IBN frequency is about 19.8~GHz and with a 50~dB S/N ratio in a 1~kHz bandwidth. The IBN signal is mixed with a synthesized RF and down-converted to the MHz range, it is then amplified and sent to a PLL electronics working with a 30~MHz local oscillator (LO). The PLL output signal is fed back to the QCL current driver for the active stabilization of the IBN phase with a 20~kHz bandwidth. Both the synthesizer and the LO are linked to the 10~MHz frequency standard provided by a GPS-disciplined Rb-locked quartz oscillator chain. In these conditions, when the PLL is active, the IBN phase is locked to that of the LO, whose frequency is stable at a level of $6 \cdot 10^{-13}$ in 1~s. In this way the phase jitter depending on the spacing fluctuations of the THz QCL-comb is suppressed. 

The THz QCL-comb offset fluctuations can be subtracted from the multi-heterodyne signal by implementing the passive scheme described in fig.~\ref{fig:THzQCLlock}-right. The multi-heterodyne signal detected by the HEB is split into two parts. The first part is filtered with a narrow band pass (BP) filter, in order to select a single BN, to be used as reference (RBN). The RBN is then shifted by 40 MHz and is mixed again with the multi-heterodyne signal. The difference-frequency component of the mixed signal is finally acquired by the real-time FFT spectrum analyzer. In this difference process the common-mode noise coming from offset fluctuations, and carried by the RBN, is automatically canceled out from all the modes.

\subsection{Relation between the FC modes and the corresponding BNs}
\label{sec:BNs_phases}

In this section we will show how the Fourier phases of the BNs are related to the Fourier phases of the FC modes that generate them. 

We start by writing the time-dependent overall electric field emitted by the two FCs involved in the dual-comb multi-heterodyne detection. 
\begin{equation}
E_\text{sample-FC} = \sum_n E_{n\text{,sample}} \exp{\imath [ 2 \pi (n f_\text{s,sample} + f_\text{o,sample}) t + \phi_{n\text{,sample}} ] } 
\label{eq:samplefield}
\end{equation}
\begin{equation}
E_\text{LO-FC} = \sum_m E_{m\text{,LO}} \exp{\imath [ 2 \pi (m f_\text{s,LO} + f_\text{o,LO}) t + \phi_{m\text{,LO}} ] } 
\label{eq:LOfield}
\end{equation}
Here $n$ and $m$ are the integers numbering the FC modes, $E_n$ is the (real) amplitude of the $n$-th mode, $f_\text{s}$ is the mode spacing, $f_\text{o}$ is the offset frequency, $t$ is time, and $\phi_{n}$ the Fourier phase of the $n$-th mode. \\ 
The intensity of the field on the detector is given by 
\begin{equation}
I = \frac{c~\epsilon_0}{2} | E_\text{sample-FC} + E_\text{LO-FC}|^2  \\  
\label{eq:dual_intensity_impl}
\end{equation}
and explicitly 
\begin{equation}
\begin{aligned}
I = & \frac{c~\epsilon_0}{2} \Bigg\{ \sum_n E_{n\text{,sample}}^2 + 2 \sum_{n>m} E_{n\text{,sample}} E_{m\text{,sample}} \exp{\imath [ 2 \pi (n-m) f_\text{s,sample}~t + \phi_{n\text{,sample}} - \phi_{m\text{,sample}} ] }   \\ 
 & \qquad + \sum_n E_{n\text{,LO}}^2 + 2 \sum_{n>m} E_{n\text{,LO}} E_{m\text{,LO}} \exp{\imath [ 2 \pi (n-m) f_\text{s,LO}~t + \phi_{n\text{,LO}} - \phi_{m\text{,LO}} ] }  \\ 
 & \qquad + \underbrace{ 2 \Re \left[ \sum_{n,m} E_{n\text{,sample}} E_{m\text{,LO}} \exp{\imath [ 2 \pi (n f_\text{s,sample} - m f_\text{s,LO} + f_\text{o,sample} - f_\text{o,LO}) t + \phi_{n\text{,sample}} - \phi_{m\text{,LO}} ] } \right] }_{\text{dual-comb BNs}} \Bigg\}
\label{eq:dual_intensity_expl}
\end{aligned}
\end{equation}
where $\epsilon_0$ is the vacuum permittivity and $c$ is the speed of light in vacuum. 

The first two lines in eq.~\ref{eq:dual_intensity_expl} represent the total intensities and the harmonics of $f_\text{s}$ of the two FCs. These terms are generally deliberately excluded by using AC-coupled detectors with a bandwidth $BW < f_\text{s,LO} ( < f_\text{s,sample} )$. \\ 
The oscillating term at line three of eq.~\ref{eq:dual_intensity_expl} describes the generated dual-comb BNs. In particular, the amplitude, the frequency and the phase of each BN are the following: 
\begin{subequations}
\label{eq:BNs_pars}
\begin{align}
A_{i\text{,BN}} & = c~\epsilon_0~E_{n\text{,sample}} E_{m\text{,LO}}   \label{eq:BN_amp}  \\
f_{i\text{,BN}} & = | n f_\text{s,sample} - m f_\text{s,LO} + f_\text{o,sample} - f_\text{o,LO} |  \label{eq:BN_freq}  \\
\phi_{i\text{,BN}} & = \phi_{n\text{,sample}} - \phi_{m\text{,LO}}  \label{eq:BN_phase} 
\end{align}
\end{subequations}
Only the BNs whose frequency falls within the acquisition bandwidth can be detected. A proper choice of the bandwidth of the optical band-pass filter, of the detection bandwidth (usually $BW < f_\text{s,LO}/2 $) and of $f_\text{s,BNs}$ (see eq.~\ref{eq:RFspacing}) ensures a one-by-one mapping of the sample FC modes in terms of RF BNs (RF-FC), considering the LO-FC as reference (see fig.~\ref{fig:dual-comb_setup} in the main text). In particular, $\phi_{i\text{,BN}}$ directly depend on the Fourier phase of the sample FC mode of interest.

\subsection{Pulse operation vs scattering amplitude of the phases}
\label{sec:phases_simulation}

\begin{figure}[!htbp]
\begin{center}
\includegraphics[width=0.8\textwidth]{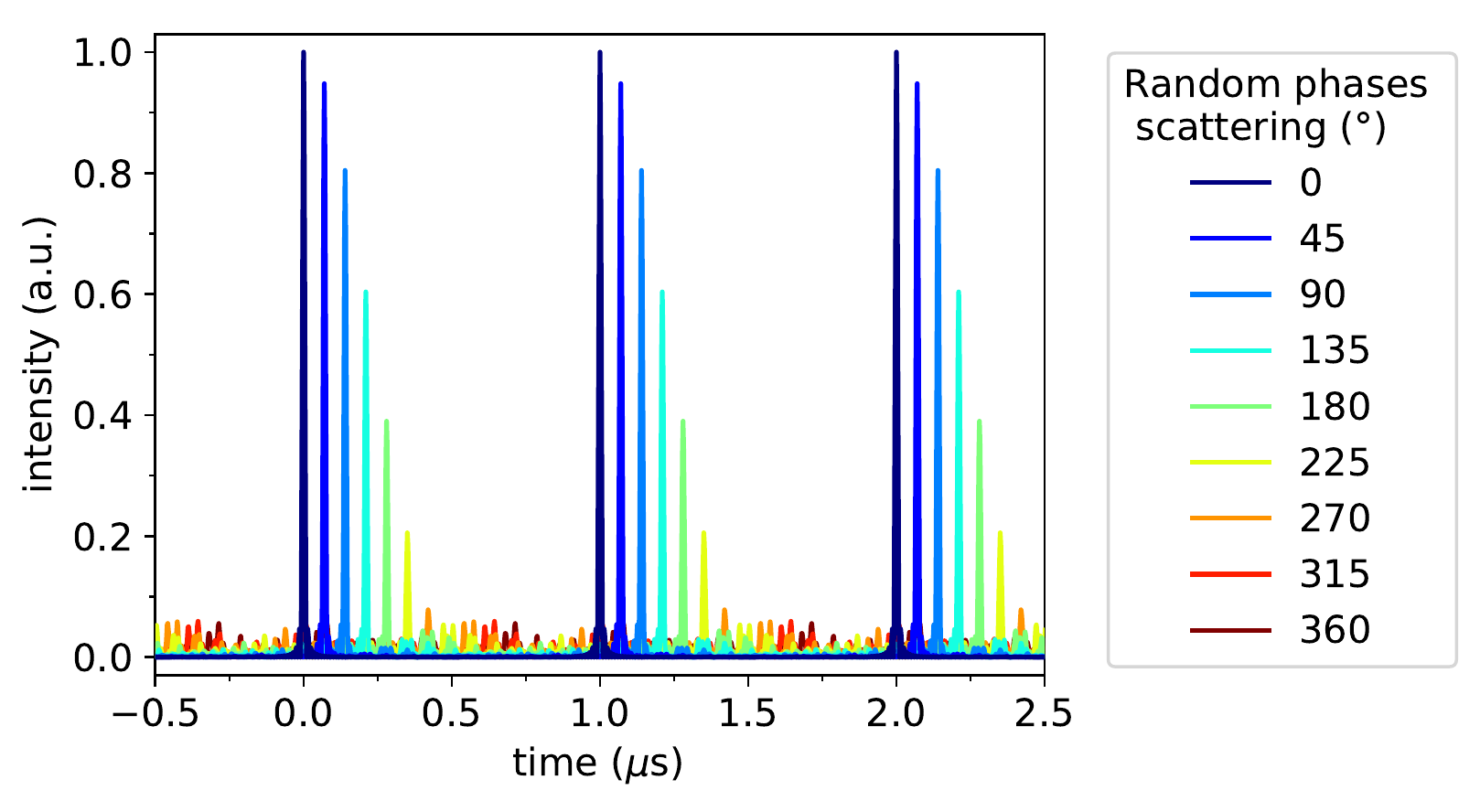}
\caption[]{Simulation of the intensity profile of a FC. \newline Temporal emission profiles obtained by varying the scattering of the random phases values. The FC is centered at 250~MHz, with $f_\text{s} = 1$~MHz and 100 modes. On the time axis consecutive curves are deliberately shifted for the sake of clarity. }
\label{fig:phases_simul_rand} 
\end{center}
\end{figure}

In this section the results of a simulation of the pulse operation of a FC are reported. In particular, the effect of the varying scattering amplitude of a random distribution of the modes phases is investigated. For this purpose, a FC centered at 250~MHz, with $f_\text{s} = 1$~MHz and emitting 100 modes, is considered. All the modes are characterized by the same intensity, while the phases distribution changes from run to run. The series of random numbers (from 0 to 1) used for generating the phases is generated only once, while the multiplying factor is changed from run to run (from $0 \degree$ to $360 \degree$). In fig.~\ref{fig:phases_simul_rand} the obtained intensity profiles are shown. 
For low scattering amplitude values the pulses are evident, while with increasing scattering amplitude the pulse--background contrast drastically degrades. Changing the series of random numbers used for generating the phases does not significantly change the attained intensity profiles.

\end{document}